\documentclass[conference, 10pt]{IEEEtran}
\renewcommand{\baselinestretch}{0.98}
\ifCLASSINFOpdf
\else
\fi
\usepackage{graphicx}
\usepackage{placeins}
\usepackage{float}
\usepackage{amssymb}
\usepackage{amsmath}
\usepackage{citesort}
\usepackage{color}
\begin{document}
\title{Two-Part Reconstruction in Compressed Sensing}

\author{\IEEEauthorblockN{Yanting Ma,\IEEEauthorrefmark{1}
Dror Baron,\IEEEauthorrefmark{1} and
Deanna Needell\IEEEauthorrefmark{2}}
\IEEEauthorblockA{\IEEEauthorrefmark{1}Department of Electrical and Computer Engineering\\
North Carolina State University;
Raleigh, NC 27695, USA\\ 
Email: $\lbrace $yma7, barondror$\rbrace $@ncsu.edu}
\IEEEauthorblockA{\IEEEauthorrefmark{2}Department of Mathematical Sciences\\
Claremont McKenna College;
Claremont, CA 91711, USA\\
Email: dneedell@cmc.edu}
}

\maketitle

\begin{abstract}
Two-part reconstruction is a framework for signal recovery in compressed sensing (CS), in which the advantages of two different algorithms are combined. Our framework allows to accelerate the reconstruction procedure without compromising the reconstruction quality. To illustrate the efficacy of our two-part approach, we extend the author's previous Sudocodes algorithm and make it robust to measurement noise. In a 1-bit CS setting, promising numerical results indicate that our algorithm offers both a reduction in run-time and improvement in reconstruction quality.
\end{abstract}

\begin{keywords}
compressed sensing, fast algorithms, two-part reconstruction.
\end{keywords}

%

\section{Introduction}
\label{sec_intro}
In the compressed sensing (CS) signal acquisition paradigm, sparse signals $x^N\in \mathbb{R}^N$ containing only $K\ll N$ nonzero coefficients can be reconstructed from measurements $y\in \mathbb{R}^M$ with $K<M\ll N$~\cite{DonohoCS,CandesRUP}. The measurement system is often modeled as a linear matrix-vector multiplication $y=\Phi x$; measurement noise can also be supported, $y=\Phi x+z$. While reconstruction quality is an important criterion for designing reconstruction algorithms, the run-time is also of great concern in practical applications.

{\bf Prior art:} There is a vast literature on CS signal reconstruction algorithms; many existing algorithms can be classified as combinatorial or geometric. The combinatorial approach uses sparse and often binary measurement matrices~\cite{Indyk2008,Iwen2013sparsematrice}, and features fast recovery but requires a suboptimal  number of measurements. Sparse binary measurement matrices based on expander graphs have been shown to have good properties for compressed sensing reconstruction problems~\cite{jafarpour2009}. The geometric appoach uses dense measurement matrices that satisfy the Restricted Isometry Property (RIP)~\cite{CandesRUP}. Examples of the geometric approach include CoSaMP~\cite{Cosamp08} and IHT~\cite{BlumensathDavies2009}. The advantages of the geometric approach are that it requires a small number of measurements and offers resiliency to measurement noise at the expense of greater run-time~\cite{Berinde2008combining}.  

The Sudocodes algorithm~\cite{sudo_isit} provides a new scheme for lossless reconstruction of sparse signals in the case where measurements are noiseless. The algorithm has two parts. In Part~1, the Sudocodes algorithm~\cite{sudo_isit} uses a sparse binary random matrix with $L$ ones in each row to acquire the measurements. Therefore, only $L$ summation operations are needed for the acquisition of each measurement. With these measurements, Part~1 can efficiently recover most of the zero coefficients and some of the nonzero coefficients in $x$. After Part~1, only a modest number of coefficients are unknown, and the unknown coefficients are solved in Part~2, where a dense measurement matrix is used. Part~2 first updates the components in the measurements and measurement matrix that are related to the coefficients recovered in Part~1. Therefore, Part~2 only needs a modest number of measurements and it applies matrix inversion to solve the remaining reconstruction problem. A variation of the Sudocodes algorithm is group testing basis pursuit CS (GBCS)~\cite{Talari2011gbcs}, which applies a CS reconstruction algorithm, Basis Pursuit, in Part~2. Sudocodes and GBCS are both fast. However, they can only be applied to the noiseless case, which is impractical in many real life applications. Nonetheless, the idea of two-part reconstruction motivates a more practical framework, which performs fast reconstruction in the presence of noise. Unlike Sudocodes, a more straightforward approach to two-part reconstruction is to perform support recovery in Part~1~\cite{gopi2013}, but exact support recovery is ambitious, especially when the measurements are noisy.

{\bf Contributions:} First, we propose a two-part framework for reconstruction of sparse signals (Section~\ref{subsec_framework}). The purpose of our framework is to accelerate the reconstruction procedure without compromising the reconstruction quality. Our strategy is to let Part~1 perform a simple algorithm to provide partial reconstruction and let Part~2 complete the residual reconstruction problem. Second, to illustrate the efficacy of our two-part approach, we extend the Sudocodes algorithm~\cite{sudo_isit} and make it robust to measurement noise  (Section~\ref{subsec_noisysudo}); we call this algorithm Noisy-Sudocodes. Third, we apply Noisy-Sudocodes to 1-bit CS, by using a modified 1-bit quantizer in Part~1 and Binary Iterative Hard Thresholding (BIHT)~\cite{Jacques2011robust} in Part~2  (Section~\ref{subsec_1bit}). Promising numerical results (Section~\ref{sec_numerical}) indicate that our algorithm offers both a reduction in run-time and improvement in reconstruction quality.

\section{Two-part reconstruction}
\subsection{Framework}
\label{subsec_framework}
\begin{figure}[ht]
\center
\noindent
  \includegraphics[width=85mm]{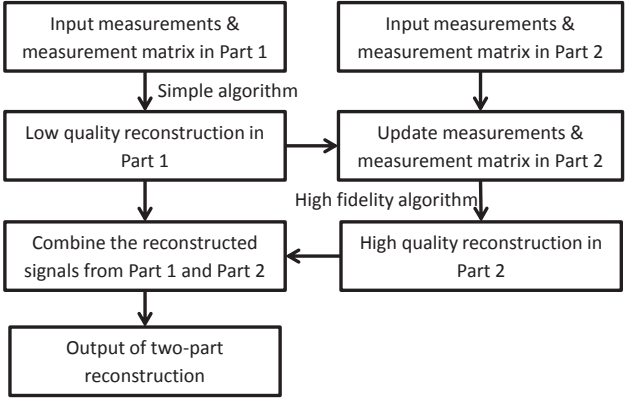}
  \caption{Block diagram for two-part reconstruction.}\label{blockdiagram}
\end{figure}

We discuss our two-part reconstruction framework, which is illustrated in Figure~\ref{blockdiagram}.
Part~1 will apply a simple and fast algorithm. This algorithm will quickly provide a low quality reconstruction in the sense that for some portion of the coefficients, it may not be able to perform sufficiently accurate reconstruction. The index set of the coefficients that are not accurately reconstructed in Part~1 will then be sent to Part~2. On the one hand, in order to reduce the run-time, we want the portion left for Part~2 to be as small as possible, because the algorithm in Part~2 is in general more complex and slower than the algorithm in Part~1. On the other hand, we don't want to sacrifice too much accuracy. The trade-off between reconstruction quality and run-time can be adjusted according to the specifics of the application at hand.

Similar to the original Sudocodes algorithm~\cite{sudo_isit}, Part~2 will only deal with the coefficients that are left over from Part~1. Part~2 will first remove the redundant coefficients in $x$ whose indices are not in the index set sent by Part~1, and also remove the corresponding columns from the measurement matrix. The measurements are then updated by subtracting the contribution of the removed coefficients to the original measurements. Because the problem size is greatly reduced in Part~2, the algorithm applied in Part~2 can put emphasis on reconstruction quality rather than run-time.
\vspace{-0.5\baselineskip}
\vspace{-0.5\baselineskip}
A potential drawback of our two-part framework is that in order for Part~1 to identify most of the coefficients correctly, it might be necessary to use an increased number of measurements. Therefore, our two-part framework is mostly applicable when fast reconstruction is crucial whereas measurements are relatively cheap. 

\subsection{Noisy-Sudocodes}
\label{subsec_noisysudo}
To illustrate how two-part reconstruction can combine the advantages of two algorithms, we describe a Noisy-Sudocodes algorithm, which extends the original Sudocodes algorithm~\cite{sudo_isit} by making it robust to measurement noise, while retaining the high-speed processing of the original algorithm.  

We begin with some notations. Let $x$ be the real-valued input signal and let $x(i)$ represent the $i$th element of $x$. Denote the measurement matrices in Part~1 and Part~2 by $\Phi _1 \in \mathbb{R}^{M_1\times N}$ and $\Phi _2\in \mathbb{R}^{M_2\times N}$, respectively. Let $z_1$ and $z_2$ represent additive measurement noise; the noisy measurements in the two parts are given by:
\begin{equation}
y_1=\Phi _1 x+z_1,
\label{eq_CSy1}
\end{equation}
\begin{equation}
y_2=\Phi _2 x+z_2.
\label{eq_CSy2}
\end{equation}
The reconstructed signal obtained from Part~1 is denoted by $\widehat{x}_1$.
Note that not all the coefficients are recovered in Part~1, because it is a simple low quality algorithm. The index set $T$ can now be defined as:
\[T\triangleq \lbrace i\text{: } x(i) \text{ is not recovered in Part~1} \rbrace.\]
Define $(x)_T \triangleq \lbrace x(i) \in x\text{: }i\in T \rbrace$. Let $(\Phi )_T$ denote the submatrix formed by combining columns of $\Phi$ at column indices $T$. We define $\Omega _r(j)$ as the support (indices of nonzeros) of the $j$th row of $\Phi _1$, and $\Omega _c(i)$ as the support of the $i$th column of $\Phi _1$, where $j\in \{1,...,M_1\}$ and $i\in \{1,...,N\}$.

The Noisy-Sudocodes algorithm proceeds as follows:

{\bf Part~1:} The measurement matrix $\Phi _1$ has independent and identically distributed (i.i.d.) Bernoulli entries. The measurement vector $y_1$ is acquired via (\ref{eq_CSy1}), and each $y_1(j)$ is the summation of a subset of coefficients of $x$ that depend on $\Omega _r(j)$. If there is no measurement noise, as in the Sudocodes algorithm~\cite{sudo_isit}, then for a real-valued input $x$, a zero measurement can only be the summation of zero coefficients. In other words, if $y_1(j)$ is zero, then $(x)_{\Omega _r(j)}= \mathbf{0}$.  But in the presence of noise, a measurement is (very) unlikely to be precisely zero. Moreover, a small-valued measurement could have measured a combination of multiple large-valued coefficients, though with small probability $p$. However, it is unlikely that a large-valued coefficient could appear in multiple small-valued measurements (if $p$ is small, then $p^n$ decreases quickly as $n$ increases). 

Our numerical experiments suggest that if the measurement matrix is sparse enough, then it is sufficiently accurate to identify a coefficient to be zero when it is involved in three or more small-valued measurements. To utilize this numerical observation, let $\epsilon$ be a small positive constant that depends on the noise level. Define an index set that contains the indices of small-valued measurements as:
\begin{equation}
\label{eq_S}
S \triangleq \lbrace j: |y_1 (j)|<\epsilon, j\in \{1,...,M_1\} \rbrace .
\end{equation}
We identify $x(i), i\in \{1,...,N\}$ to be zero if $|\Omega _c(i)\cap S|\geq 3$, where $| \cdot |$ denotes cardinality. For those coefficients that cannot be identified in this zero-identification procedure, the indices are recorded in $T$ and sent to Part~2.

{\bf Part~2:} Solve the remaining reconstruction problem by utilizing a high quality CS reconstruction algorithm. The distribution of the measurement matrix $\Phi _2$ depends on the algorithm applied in Part~2 (for example, if CoSaMP~\cite{Cosamp08} is used in Part~2, then a Gaussian $\Phi _2$ is appropriate). Initially, the measurement vector $y_2$ is acquired via (\ref{eq_CSy2}).  After receiving $T$ from Part~1, Part~2 first updates $x$ and $\Phi _2$: 
\begin{equation*}
\widetilde{x} = (x)_T,
\end{equation*}
\begin{equation*}
\widetilde{\Phi}_2 = (\Phi _{2})_{T}.
\end{equation*}
Note that we only identify zero coefficients in Part~1. The zero coefficients do not contribute to $y_2$, thus $y_2$ need not be updated. The high quality CS reconstruction algorithm in Part~2 takes $\widetilde{x}$, $\widetilde{\Phi}_2$, and $y_2$, and computes $\widehat{x}_2$, the reconstructed signal.

We complete the reconstruction by assigning the coefficients in $\widehat{x}_2$ at indices $T$ to the elements in the final reconstructed signal $\widehat{x}$,
\begin{equation*}
(\widehat{x})_{T} = \widehat{x}_2. 
\end{equation*}

\subsection{Application to 1-bit compressed sensing}
\label{subsec_1bit}
In 1-bit CS~\cite{Boufounos2008}, the CS measurements are quantized to 1 bit per measurement. The problem model for noiseless and noisy 1-bit CS is formulated as
\begin{equation}
y = \text{sign} (\Phi x),
\label{eq_noiseless 1 bit}
\end{equation} 
\begin{equation}
y = \text{sign} (\Phi x+z),
\label{eq_noisy 1 bit}
\end{equation}
where $z$ is measurement noise. Note that the measurements acquired in both noiseless and noisy 1-bit CS include quantization noise. 
The quantization noise explains why the SNR achieved in the noiseless 1-bit CS setting, which is shown in Figure \ref{noiseless}, is finite, whereas unquantized noiseless measurements yield perfect reconstruction~\cite{DonohoCS,CandesRUP}.

Because the amplitude information of the measurements is lost due to the quantization described by (\ref{eq_noiseless 1 bit}) or (\ref{eq_noisy 1 bit}), it is convenient to assume that the 1-bit CS framework imposes a unit energy constraint on the reconstructed signal.

In Part~1 of Noisy-Sudocodes discussed in Subsection~\ref{subsec_noisysudo}, we only need to know if $y_1(j)$ is greater or less than $\epsilon$. Therefore, 1 bit is sufficient to quantize each measurement without losing any information needed for the reconstruction in Part~1. For example, we can quantize $y_1(j)$ as:

\begin{equation}
\label{eq_quantize y}
\bar{y}_1(j)= 
\begin{cases} 
0, &\text{if } |y_1(j)|\le \epsilon\\
1, &\text{if } |y_1(j)|> \epsilon
\end{cases}.
\end{equation}
We note that this modified 1-bit quantizer~(\ref{eq_quantize y}) is only used in Part~1, whereas in Part~2 we utilize a standard 1-bit quantizer~(\ref{eq_noiseless 1 bit}, \ref{eq_noisy 1 bit}).

Then utilizing $\bar{y_1}$ as the measurements, (\ref{eq_S}) can be rewritten as:
\begin{equation*}
\label{eq_Snew}
S \triangleq \lbrace j: |\bar{y}_1(j)| = 0, j\in \{1,...,M_1\} \rbrace.
\end{equation*}
This discussion implies that Noisy-Sudocodes can be extended to a 1-bit CS setting by utilizing a 1-bit CS algorithm in Part~2.

A possible 1-bit CS algorithm that can be utilized is BIHT~\cite{Jacques2011robust}. BIHT achieves better reconstruction performance than the previous 1-bit CS algorithms in the noiseless 1-bit CS setting. We show by numerical results in Section~\ref{sec_numerical} that combining Noisy-Sudocodes with BIHT in a two-part setting (Sudo+BIHT) achieves better reconstruction quality and reduction in run-time than directly using BIHT (direct BIHT). 

\section{Numerical results}
\label{sec_numerical}
In this section, we present simulation results that compare Sudo+BIHT and direct BIHT in terms of SNR and run-time. SNR is defined as
\begin{equation*}
\text{SNR(dB)}\triangleq 10\log _{10}(\| x\| _2^2/\|x-\widehat{x}\| _2^2),
\end{equation*}
where $x$ is the input signal and $\widehat{x}$ is the reconstructed signal; run-time is measured in seconds on a Dell OPTIPLEX 9010 running an Intel(R) $\text{Core}^{\text{TM}}$ i7-3770 with 16GB RAM.

We simulate both noiseless, in which BIHT-$l1$ is utilized and noisy 1-bit CS settings, in which BIHT-$l2$ is utilized. The input signal $x$ is of length $N=10,000$, containing $K=50$ nonzero coefficients, which are i.i.d. Gaussian with zero mean, and $x$ is normalized such that $\| x\| _2 = 1$. Let $M_1$ and $M_2$ be the number of measurements for Parts~1 and~2 of Sudo+BIHT. Then $M=M_1+M_2$ is the number of measurements for direct BIHT. We perform the trials for measurement rate $M/N$ within the range $[0,2]$. In our simulation, we let $M_1 = c_1K\log _2(N/K)$, which for sufficiently large $c_1$ (determined numerically) allows Part~1 to identify more than $90\%$ of the zero coefficients. The measurement matrix $\Phi _1 \in \mathbb{R}^{M_1\times N}$ is i.i.d. Bernoulli with Bernoulli parameter $p=\frac{c_2}{K}$, where $c_2$ (determined numerically) is a constant. Note that the nonzero entries of the Bernoulli matrix are scaled by $\frac{1}{\sqrt{p}}$ in order to have the same input SNR as in direct BIHT. $\Phi _2 \in \mathbb{R}^{M_2 \times N}$ is i.i.d. Gaussian with $\phi _2(i,j) \sim  \mathcal{N}(0,1)$, $i\in \{1,...,M_2\}\text{, } j\in \{1,...,N\}$.  

For direct BIHT, the measurement matrix $\Phi \in \mathbb{R}^{M\times N}$ is i.i.d. Gaussian with $\phi (i,j)\sim  \mathcal{N}(0,1)$, $i\in \{1,...,M\}\text{, } j\in \{1,...,N\}$.

Finally, the additive measurement noise $z$, which we use in the noisy setting, is i.i.d. Gaussian. It has zero mean and its variance is $10^{-2.5}$.

{\bf Noiseless setting:} The measurement vector $y_1$ for Part~1 of Sudo+BIHT is acquired by (\ref{eq_quantize y}) with $\epsilon = 0$, and the measurement vectors $y_2$ for Part~2 of Sudo+BIHT and $y$ for direct BIHT  are acquired  by (\ref{eq_noiseless 1 bit}). In the noiseless setting, if any element $y_1(j)$ only measures zero coefficients, then $y_1(j)$ will be strictly zero. Therefore, we modify Part~1 of Noisy-Sudocodes by identifying $x(i)$ to be zero if it is measured at least once in the zero measurements, i.e., $|\Omega _c(i)\cap S|\geq 1$. Note that in this case, Part~1 will not introduce any error. 
\vspace{-0.5\baselineskip}
\begin{figure}[ht]
\centering
\noindent
  \includegraphics[width=95mm]{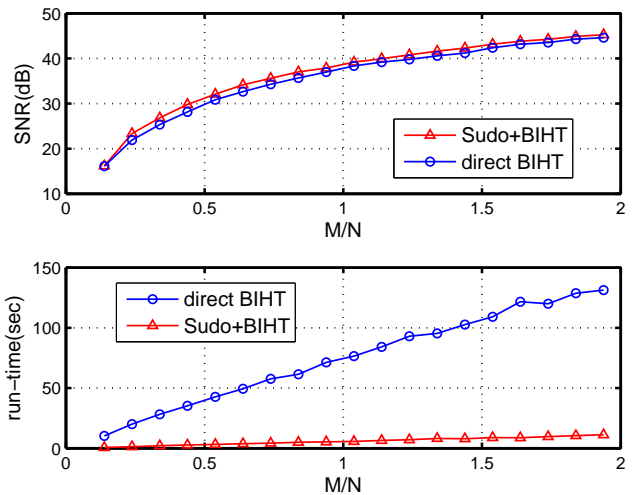}
  \caption{Reconstruction performance from 1-bit measurements in the noiseless setting.}\label{noiseless}
\end{figure}
The simulation results for the SNR and run-time are shown in Figure \ref{noiseless}. We iterate over BIHT until the consistency property\footnote{We say that the consistency property of BIHT~\cite{Boufounos2008} is satisfied if applying the measurement and quantization system (\ref{eq_noiseless 1 bit}) and (\ref{eq_noisy 1 bit}) to the reconstructed signal $\widehat{x}$ yields the same measurements $y$ as the original measurements.} is satisfied or the number of iterations reaches 100. We notice that Sudo+BIHT achieves slightly higher SNR than direct BIHT except in the low measurement rate ($M/N$) region, where the SNR for both Sudo+BIHT and direct BIHT is modest. Owing to the generally low reconstruction quality in the low measurement rate region, it is more interesting to compare performance in the higher measurement rate region. It is demonstrated in Figure \ref{noiseless} that as $M/N$ increases, the SNR for both algorithms increases similarly. However, the run-time for Sudo+BIHT grows slower than direct BIHT. 

{\bf Noisy setting:} The measurement vector $y_1$ for Part~1 of Sudo+BIHT is acquired by (\ref{eq_quantize y}) with $\epsilon > 0$, and the measurement vectors for Part~2 of Sudo+BIHT, $y_2$, and direct BIHT, $y$, are acquired  by (\ref{eq_noisy 1 bit}).
\vspace{-0.5\baselineskip}
\begin{figure}[ht]
\centering
\noindent
  \includegraphics[width=95mm]{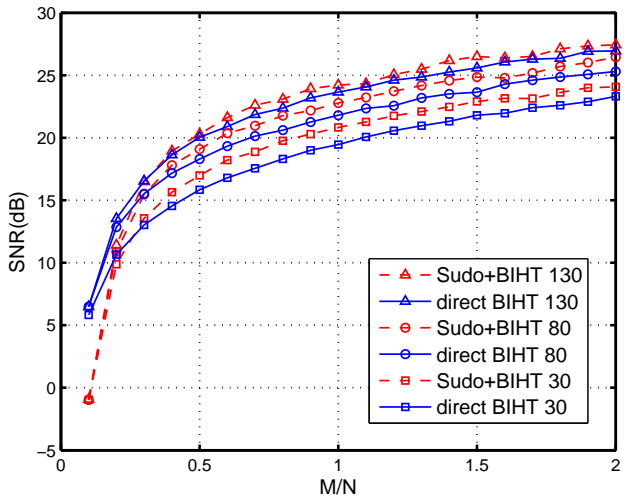}
  \caption{SNR achieved by Sudo+BIHT and direct BIHT in the noisy 1-bit CS setting with 30, 80 and 130 iterations for BIHT.}\label{mixSNR}
\end{figure}
\vspace{-0.5\baselineskip}
\begin{figure}[ht]
\centering
\noindent
  \includegraphics[width=95mm]{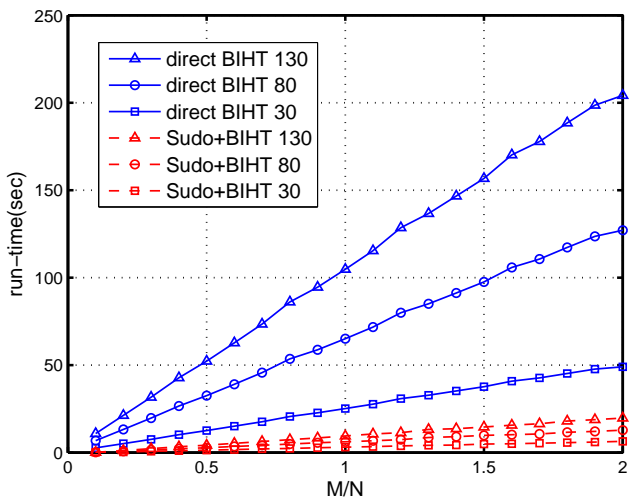}
  \caption{Run-time of Sudo+BIHT and direct BIHT in the noisy 1-bit CS setting with 30, 80 and 130 iterations for BIHT.}\label{mixTIME}
\end{figure}
\vspace{-0.5\baselineskip}
The resulting SNR is shown in Figure~\ref{mixSNR}, and run-time is shown in Figure~\ref{mixTIME}. When the number of iterations for BIHT is 30 in both Part~2 of Sudo+BIHT and direct BIHT, Sudo+BIHT yields better consistency and thus provides better reconstruction quality. With more iterations, the SNR for both Sudo+BIHT and direct BIHT improves. The SNR curve of direct BIHT tends to get closer to Sudo+BIHT as the number of iterations increases, because for Sudo+BIHT, the error introduced in Part 1 cannot be corrected by Part 2. We notice that the run-time for Sudo+BIHT with 130 BIHT iterations is half of that for direct BIHT with 30 BIHT iterations, while the SNR increased by roughly 5 dB. In other words, problem size reduction due to zero identification in Part~1 allows BIHT in Part~2 to run more iterations to improve reconstruction quality with reasonable run-time. 

\section{conclusion}
\label{sec_conclusion}
We discussed a two-part framework for fast reconstruction of sparse signals, in which Part~1 quickly reduces the problem size by reconstructing the ``easy" part, leaving a ``difficult" problem of smaller size for Part~2.
The zero-identification algorithm in Noisy-Sudocodes is well suited for Part~1 of our two-part framework, because it is fast. Part~1 of Noisy-Sudocodes quickly identifies most of the zero coefficients without introducing much error. Therefore, a high fidelity algorithm in Part~2 is able to complete the reconstruction efficiently due to the reduction in problem size.
The promising simulation results of Noisy-Sudocodes with BIHT in Part~2 (Sudo+BIHT) implies that Noisy-Sudocodes could be promising for algorithm design in 1-bit CS reconstruction problems. 

\bibliographystyle{IEEEtran}
\bibliography{cites}
\end{document}